\documentclass[a4paper,fleqn,usenatbib]{mnras}

\usepackage{newtxtext,newtxmath}
\usepackage[T1]{fontenc}
\usepackage{ae,aecompl}

\usepackage{graphicx}	% Including figure files
\usepackage{amsmath}	% Advanced maths commands
\usepackage{amssymb}	% Extra maths symbols

\newcommand{\speed}[1]{#1 km~s${}^{-1}$}
\newcommand{\accel}[1]{#1 km~s${}^{-2}$}
\newcommand{\accell}[1]{#1 m~s${}^{-2}$}
\newcommand{\nfig}[1]{Figure~\ref{#1}}

\title[Simultaneous observations of three types of waves in a jet event]{Coronal EUV, QFP, and kink waves simultaneously launched during the course of jet-loop interaction}

\author[Yuandeng Shen]{
Yuandeng Shen,$^{1,2,3,4}$\thanks{E-mail: ydshen@ynao.ac.cn}
Zehao Tang,$^{1,4}$
Hongbo Li,$^{1,4}$
and Yu Liu$^{1,3,4}$
\\
$^{1}$Yunnan Observatories, Chinese Academy of Sciences,  Kunming, 650216, People's Republic of China\\
$^{2}$State Key Laboratory of Space Weather, Chinese Academy of Sciences, Beijing 100190, People's Republic of China\\
$^{3}$Center for Astronomical Mega-Science, Chinese Academy of Sciences, Beijing, 100012, People's Republic of China\\
$^{4}$University of Chinese Academy of Sciences, Beijing 100049, People's Republic of China
}

\date{Accepted 2018 July 9 Received 2018 July 6; in original form 2018 May 8}

\pubyear{2018}

\begin{document}
\label{firstpage}
\pagerange{\pageref{firstpage}--\pageref{lastpage}}
\maketitle

% Abstract of the paper
\begin{abstract}
We present the observations of an extreme ultraviolet (EUV) wave, a quasi-periodic fast-propagating (QFP) magnetosonic wave, and a kink wave that were simultaneously associated with the impingement of a coronal jet upon a group of coronal loops. After the interaction, the coronal loop showed obvious kink oscillation that had a period of about 428 seconds. In the meantime, a large-scale EUV wave and a QFP wave are observed on the west of the interaction position. It is interesting that the QFP wave showed refraction effect during its passing through two strong magnetic regions. The angular extent, speed, and lifetime of the EUV (QFP) wave were about $140^\circ$ ($40^\circ$), \speed{423 (322)}, and 6 (26) minutes, respectively. It is measured that the period of the QFP wave was about $390 \pm 100$. Based on the observational analysis results, we propose that the kink wave was probably excited by the interaction of the jet; the EUV was probably launched by the sudden expansion of the loop system due to the impingement of the coronal jet; and the QFP wave was possibly formed through the dispersive evolution of the disturbance caused by the jet-loop interaction.
\end{abstract}

\begin{keywords}
activities--flares--corona--magnetic fields--waves--jets
\end{keywords}

\section{Introduction}
The magnetized coronal plasma can support the propagation of various kinds of magnetohydrodynamics (MHD) waves that are important for remote diagnosing coronal physical parameters such as the magnetic field. In addition, the dissipation of wave energy into the coronal plasma is thought to be an important energy source for heating the corona. Therefore, MHD waves in the solar atmosphere has been a main hot research subject in solar physics for many years \citep[see,][and reference therein]{2005LRSP....2....3N}.

The unambiguous imaging observations of quasi-periodic fast-propagating (QFP) magnetosonic waves were firstly reported by \cite{2011ApJ...736L..13L} using the high resolution observations taken by the Atmospheric Imaging Assembly \citep[AIA;][]{2012SoPh..275...17L} onboard the {\em Solar Dynamics Observatory} \citep[{\em SDO};][]{2012SoPh..275....3P}. \cite{2014SoPh..289.3233L} summarized that QFP wave's speed, deceleration, and period are respectively in the ranges of \speed{500--2200},  \accel{1--4}, and 25--400 second; they are often accompanied by flares, but their initial appearance positions are often in a few megametres to the flare locations. So far the driving mechanism of QFP waves is still unclear. Recent observations indicated that QFP waves often share one or more periods with the accompanying flares \citep[e.g.,][]{2011ApJ...736L..13L,2012ApJ...753...52L,2012ApJ...753...53S,2013SoPh..288..585S,2018ApJ...853....1S}, thus the nonlinear energy releasing processes in magnetic reconnections are thought to be the possible exciter \citep[e.g.,][]{2006ApJ...644L.149O,2000A&A...360..715K,2012ApJ...749...30M}. Besides, the leakage of photospheric oscillations to the corona \citep[e.g.,][]{2012ApJ...753...53S} and the dispersively evolution of an initial broad-band disturbance \citep[e.g.,][]{2014A&A...569A..12N,2018MNRAS.477L...6S,2018ApJ...860L...8S} are also possible drivers for QFP waves. Other observational and thoretical works have also performed to understand the excitation and evolution of QFP waves, as well as their applications in coronal seismology \citep[e.g.,][]{2011ApJ...740L..33O,2013A&A...554A.144Y,2013A&A...560A..97P,2015ApJ...800..111Y,2016A&A...594A..96G,2016ApJ...823..150T,2017ApJ...844..149K,2017ApJ...851...41Q,2018ApJ...860...54O}.

Large-scale extreme ultraviolet (EUV) waves have been studied for twenty years \citep{1998GeoRL..25.2465T}. In earlier studies, the main discrepancies about EUV waves are about their driving mechanism and physical nature. Some authors consider that EUV waves are driven by flare pressure pulses \citep[e.g.,][]{2004AA...418.1117W,1999SoPh..187...89C}, while others proposed that they are driven by coronal mass ejections (CMEs) \citep[e.g.,][]{2006ApJ...641L.153C,2011ApJ...738..160M,2012ApJ...752L..23S}. For the physical nature, observations indicated that EUV waves showed both wave and non-wave properties. In recent years, more and more observational studies based on high resolution observations suggested that EUV waves should be CME driven fast-mode magnetosonic waves or shocks. This scenario has been confirmed by many observations and theoretical studies of the separation process between CMEs and waves \citep[e.g.,][]{2008ApJ...680L..81L,2009ApJ...691L.123G,2012ApJ...756..143O,2013ApJ...775...39Y,2013ApJ...773L..33S,2017ApJ...851..101S}. In addition, EUV waves can also be driven by sudden loop expansions caused by external disturbances \citep{2018ApJ...860L...8S} and coronal jets \citep{2018arXiv180512303S}. For more details of EUV waves, one can refer to \cite{2014SoPh..289.3233L} and \cite{2015LRSP...12....3W}.

Here, we report the observations of an EUV, a QFP, and a kink wave that were  simultaneously associated with the impingement of a coronal jet upon a group of coronal loop system. The AIA images and the line-of-sight (LOS) magnetograms taken by the Helioseismic and Magnetic Imager \citep[HMI;][]{2012SoPh..275..327S} onboard the {\em SDO} have a pixel size of $0\arcsec.6$, and the cadences of the AIA and HMI magnetograms are 12 and 45 seconds, respectively. Next section presents the main observational results, discussions and conclusions are given in the last section.

\section{Results}
\begin{figure}
\includegraphics[width=0.9\columnwidth]{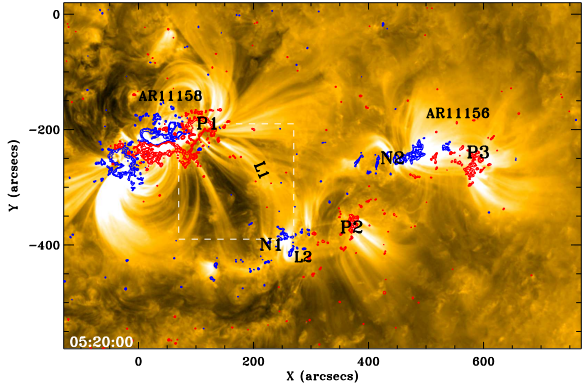}
\caption{An AIA 171 \AA\ direct image overlaid with the contours of the HMI LOS magentogram at 05:20:42 UT, in which the red and blue contours represent the regions of positive and negative polarities, respectively. Notations P1, P2, and P3 (N1 and N2) indicate the positive (negative) magnetic polarities.}
\label{fig1}
\end{figure}

The event occurred on 2011 February 14 in the NOAA active region AR11158. The pre-eruption magnetic configuration is displayed in \nfig{fig1}, in which the red and blue contours respectively indicate the positive and negative magnetic polarities. In addition, the positive polarities are marked with letters P1, P2, and P3, while the negative polarities are marked with N1 and N2. In \nfig{fig1}, one can observe a large interconnecting loop system (L1) that connected P1 in AR11158 and N1 in AR11156, and another loop (L2) system that connected N1 and P2 in AR11156 .

\begin{figure}
\includegraphics[width=0.9\columnwidth]{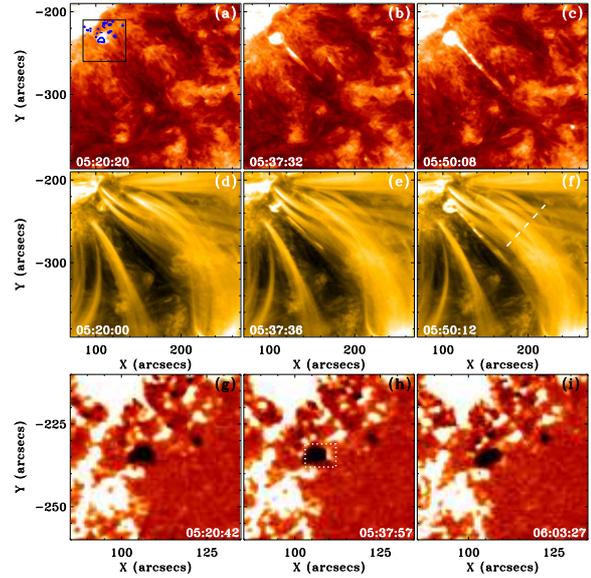}
\caption{The close up view of the box region as shown in \nfig{fig1}. The top and middle rows are AIA 304 \AA\ and 171 \AA\ images, while the bottom row shows the HMI LOS magnetogram within the box region as shown in panel (a). The white and black patches in the magnetograms represent positive and negative magnetic polarities, respectively. The magnetograms are scaled to the range from -100 to 100 Gauss.}
\label{fig2}
\end{figure}

\nfig{fig2} shows the eruption of the jet and the evolution of the magnetic field in the eruption source region. One can see a small bright point on the south of P1 (see \nfig{fig2} (a) and (d)), which became brighter from about 05:22:00 UT; and a collimated bright jet erupted from it at about 05:26:44 UT at a speed of about \speed{195}; and it reached the maximum length at about 05:50:00 UT. The jet directly impinged upon the southern end of L1 and caused the kink oscillation of the loop, which is consistent with the statistical results that most of kink oscillation of coronal loops are excited by lower coronal eruptions such as jets \citep{2015A&A...577A...4Z}. By checking the time evolution of the magnetic fluxes in the eruption source region, it is found that the region of the bright point in EUV observations was a small negative polarity. A small positive polarity emerged on the western side of the negative polarity at about 05:22:00 UT, then it disappeared at about 06:03:27 UT (see the bottom row of \nfig{fig2}). The variations of the positive and negative fluxes within the box region in \nfig{fig2} (h) are studied. It is found that the emergence of the positive flux started from 05:22:00 UT, consistent with with the start time of the bright point's brightening. After 05:52:00 UT, it rapidly decreased to the normal level of about $15 \times 10^{17}$ Maxwell. In the meantime, the absolute value of the negative flux started to decrease at about 05:22:00 UT. This time is consistent with the start time of the emergence of the positive flux. During the time interval 05:22:00 UT to 05:52:00 UT, it is measured that the increase (decrease) value of the positive (negative) flux was about $4 \times 10^{18}$ ($9 \times 10^{18}$) Maxwell, and the corresponding increasing (decreasing) speed was about $2.2 \times 10^{15}$ ($5.0 \times 10^{15}$) Maxwell $\rm s^{-1}$. The variation pattern of the magnetic fluxes suggests that magnetic cancellation occurred during the emerging process of the small positive polarity. The close temporal relationship between the magnetic fluxes and the jet suggests that the coronal jet was probably triggered by the flux cancellation between the small emerging positive polarity and the nearby pre-existing negative polarity, consistent with previous observational results \citep[e.g.,][]{2011ApJ...735L..43S,2012ApJ...745..164S,2017ApJ...851...67S,2018Ap&SS.363...26L}.

\begin{figure}
\includegraphics[width=0.9\columnwidth]{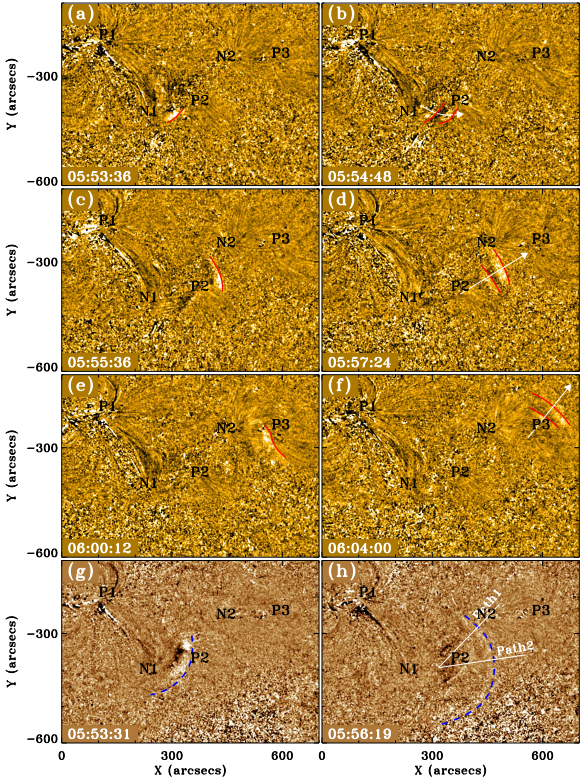}
\caption{AIA 171 \AA\ ((a) -- (f)) and 193 \AA\ ((g) -- (h)) running ratio images show the propagation of the waves, in which the wavefronts in the AIA 171 \AA\ images are highlighted with red curves, while those in the AIA 193 \AA\ images are indicated by blue dashed curves. The white arrows in panels (b), (d), and (f) indicate the propagating directions of the QFP wave. The magnetic polarities are indicated by the letters N1, N2, P1, P2, and P3, whose coordinate positions are the same as shown in \nfig{fig1}. An animation is available in the online journal of this figure.}
\label{fig4}
\end{figure}

\begin{figure}
\includegraphics[width=0.9\columnwidth]{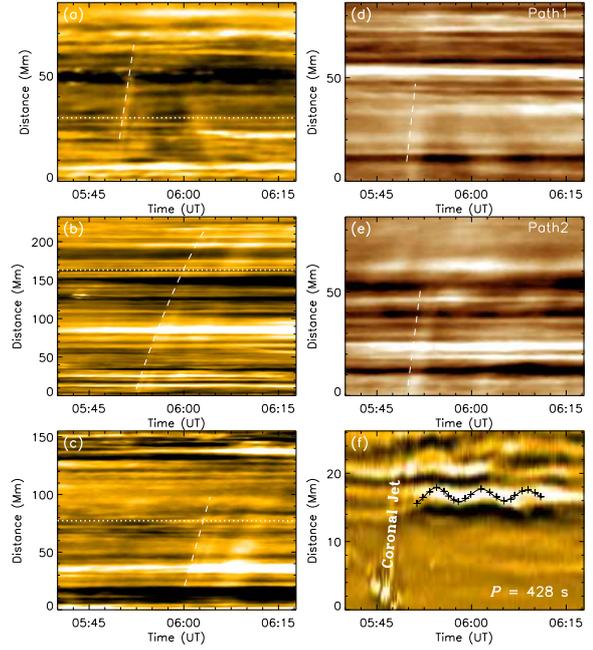}
\caption{Panels (a) -- (c) show the TDs obtained from AIA 171 \AA\ percentage images along the paths as shown by the arrows in \nfig{fig4} (b), (d), and (f), respectively. Panels (d) and (e) show the TDs obtained from AIA 193 \AA\ percentage images along path 1 and 2 as shown in \nfig{fig4} (h). The dashed lines are linear fit to the wavefronts, while the curve in panel (b) is the second order polynomial fit to the wavefront. Panel (f) shows the TD made from 171 \AA\ observations along the line as shown in \nfig{fig2} (f).}
\label{fig5}
\end{figure}

The coronal jet impinged upon L1's southern end at about 05:52:36 UT, which directly launched multiple bright wavefronts in the western side of N1 and kink oscillation of the coronal loop. The temporal evolution of the waves in the AIA 171 and 193 \AA\ running ratio images are displayed in \nfig{fig4}. Here, a running ratio image is obtained by dividing an image by the one at 36 seconds before. In the 171 \AA\ images, the wave's propagation direction changed two times when it passed P2 and P3 (see the white arrows in \nfig{fig4}). One can see that the wavefronts first propagated in west direction (see \nfig{fig4} (a) and (b)), then they changed to northwest when they passed P2 (see \nfig{fig4} (c) and (d)), and finally their propagation direction became more close to the north when they passed N2 and P3 (see \nfig{fig4} (e) and (f)). The changing propagation direction of the wavefronts indicate the occurrence of refraction effect when they passed through some macroscopic inhomogeneous coronal structures whose size are larger than the wavelength of the wavefronts. Such refraction effect was also observed in large-scale EUV waves, and the large speed gradient around the boundary of magnetic polarities are possibly important for the appearance of such a phenomenon \citep{2012ApJ...754....7S,2013ApJ...773L..33S}. In the AIA 193 \AA\ observations, the wavefronts observed in the AIA 171 \AA\ images did not appeared. However, one can observe a large-scale arc-shaped wavefront, which appeared immediately after the interaction between the jet and L1. Based on their different characteristics, the observed waves in the AIA 171 and 193 \AA\ observations can be regarded as QFP and EUV waves, repsectively. It is measured that the angular extents (lifetimes) of the QPF and EUV waves are about $40^\circ$ (26 minutes) and $140^\circ$ (6 minutes), respectively. In addition, by comparing the time evolution of the QFP and EUV waves during their initial stage (say, between N1 and P2), one can find the appearance of the EUV wave (about 05:52:55 UT) was slightly earlier than the QFP wave (05:53:24 UT); thus that they are different propagating features and should have different origins.

The kinematics of the waves and the oscillation of L1 are studied by using time-distance diagrams (TDs) and the results are shown in \nfig{fig5}. The wavefronts show as inclined bright ridges, and their slope represent the propagation speed of the waves. By applying a linear fit to the ridges, we obtained that the speeds of the QFP wave are of \speed{325, 322, and 318} during the three stages, while the average speed of the EUV wave along paths 1 and 2 are about \speed{452 and 393}, respectively. It is calculate that the average speeds of the QFP and EUV waves are \speed{322 and 423}, respectively. The propagation of the QFP waves showed obvious deceleration during the second stage. By applying a second order polynomial fit to the ridge, it is obtained that the deceleration is about \accell{138}. In \nfig{fig5} (f), one can find the oscillation of the loops after the impingement of the jet. By fitting the data points (black plus signs) with a damped vibration equation in form of $F(t) = A \exp(-\frac{t}{\tau}) \sin(\omega t+\phi)$, the oscillation parameters of the loop can be obtained. The fitting result indicates that the amplitude ($A$), period ($T$), and the damping time ($\tau$) of the loop oscillation were about 1.2 Mm, 428 second, and 36 minute, respectively.

\begin{figure}
\includegraphics[width=0.9\columnwidth]{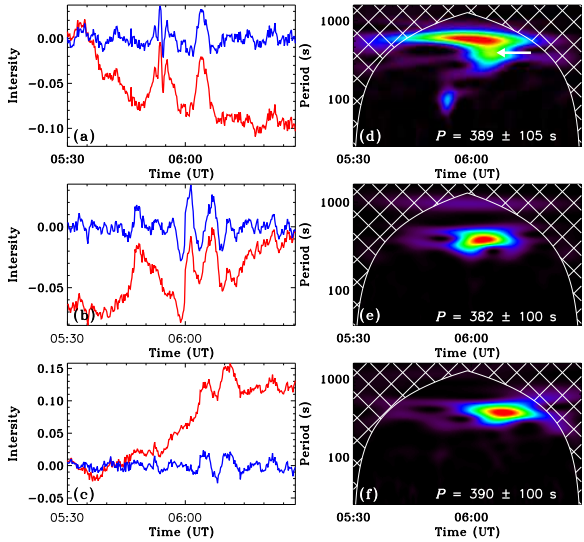}
\caption{Panels (a) -- (c) show the variations of the intensities along the horizontal lines as shown in \nfig{fig5} (a) -- (c), respectively. The red and blue curves are the percentage and detrended percentage intensities, respectively. Here, a detrended curve is obtained by subtracting the flux by the corresponding smoothed flux using a 6 minutes boxcar. For each detrended intensity curve, the corresponding wavelet power map is shown on the right, in which the white grid indicate the cone of influence region due to the edge effect of the data.}
\label{fig6}
\end{figure}

To analyze the periodicity of the QFP wave, the percentage intensities extracted from the TDs (red) along the dotted lines as shown in \nfig{fig5} (a) -- (c) are plotted in \nfig{fig6} (a) -- (c), respectively. In addition, the detrended intensities (blue) are also plotted in the figure, which show the wavefronts more clearly. The percentage intensities also indicate that the intensity variations relative to the background is about 3\%, consistent with those detected in quasi-periodic fast-propagating magnetosonic waves \citep[e.g.,][]{2011ApJ...736L..13L,2012ApJ...753...53S,2013SoPh..288..585S,2018MNRAS.477L...6S,2018ApJ...853....1S}. With the method of wavelet analysis \citep{1998BAMS...79...61T}, the periods of these intensity curves are obtained, and the results are plotted in the right column of \nfig{fig6}. The periods in the second and third stages are of $382 \pm 100$ and $390 \pm 100$ seconds, respectively. For the period in the first stage, the main period in the wavelet power map started earlier than the start time of the wavefronts, therefore, this period can not be regarded as the period of the wave. Considering the temporal relation with the intensity curve, we think that the period signal indicated by the arrow in \nfig{fig6} (d) should be caused by the wave, which indicates that the period of the wave during the first stage is $389 \pm 105$ second. Here, the periods and the corresponding errors are determined by the center values and the widths of the periodic signal in the wavelet power maps. The periods show little difference during the three stages, and they are relatively smaller than the period of the oscillating loop.

\section{Conclusions \& Discussions}
We present the simultaneous observations of an EUV, a QFP, and a kink wave in a single event occurred on 2011 February 14 in AR11158, using the high temporal and high spatial resolution observations taken by the {\em SDO}/AIA. It is observed that a coronal jet ejected from AR11158 and then it directly impinged upon the southern end of a group of interconnecting loop system that connected AR11158 and AR11156, which not only resulted in the kink oscillation of the interconnecting loop, but also an arc-shaped EUV wave and a simultaneous QFP wave in AIA 195 \AA\ and 171 \AA\ observations, respectively. The amplitude, period, and damping time of the loop oscillation were about 1.2 Mm, 428 second, and 36 minutes, respectively. It is interesting that the propagation direction of the QFP wave subjected consecutive changes from west to northwest, and this phenomenon can be regarded as the refraction effect of the QFP wave during its passing through two regions of high-intensity magnetic field \citep[see also;][]{2013ApJ...773L..33S,2018ApJ...853....1S}. 

By checking the temporal and spatial relationships between the jet and the variations of the magnetic fluxes within the eruption source region, it is found that the jet eruption was tightly related to the emergence of a small positive polarity and its cancellation with the nearby pre-existing negative polarity. It is measured that the jet speed was about \speed{195}, and the increasing (decreasing) speed of the positive (absolute value of the negative) flux was about $2.2 \times 10^{15}$ ($5.0 \times 10^{15}$) Maxwell $\rm s^{-1}$. The average speed of the QFP wave was about \speed{322}, and its propagation during the second stage showed a deceleration of about \accell{138}. Periodicity analysis of the loop oscillation and the QFP wave indicates that the period of the loop oscillation was about 428 second, and that of the QFP wave was $390 \pm 100$ seconds.

The arc-shaped EUV wave propagated at an average speed of about \speed{423}, which was faster than the QFP wave. In addition, there are still other difference between the EUV and the QFP wave, including 1) the start time of the EUV wave was earlier than the following QFP wave; 2) the angular extent of the QFP wave ($40^\circ$) was much less than the EUV wave ($140^\circ$); 3) the lifetime of the EUV wave (6 minutes) was much shorter than the QFP wave (26 minutes); and 4) their propagation distances and observing wavelengths were also different. Therefore, we propose that they are two different kinds of waves, although they were associated with the same source and observed in closed plasma temperatures.

We think that the observed waves should be excited by different physical mechanisms, but all of them were dynamically associated with the interaction between the loop system and the coronal jet. For the EUV wave, its lifetime is much shorter than typical hour-long lifetime of normal EUV waves driven by CMEs \citep[see,][and reference therein]{2014SoPh..289.3233L}. However, it is similar to the EUV waves driven by sudden loop expansions caused by the impingement of external eruptions such as jets \citep{2018ApJ...860L...8S}, or the expansion of the newly formed reconnected loops resembling of the slingshot mechanism in the eruption of coronal jets \citep[e.g.,][]{2015ApJ...804...88S}. Since no CME was associated with the present event, and considering the temporal and spatial relationship between the EUV wave and the jet, we therefore propose that the observed EUV wave was possibly excited by the sudden expansion of the interconnecting loop system caused by the impingement of the coronal jet, supporting the recent observational results presented in \cite{2018ApJ...860L...8S}.

Previous observational studies indicated that the excitation of QFP waves include several possible mechanisms, including 1) the nonlinear energy-releasing processes in the magnetic reconnection process of flares \citep[e.g.,][]{2011ApJ...736L..13L,2012ApJ...753...53S,2013SoPh..288..585S,2018ApJ...853....1S}, 2) the leakage of photospheric pressure-driven oscillations into the corona \citep[e.g.,][]{2012ApJ...753...53S}, and 3) the dispersive evolution of an initially broad-band disturbance in inhomogeneous medium \citep[e.g.,][]{2014A&A...569A..12N,2018MNRAS.477L...6S,2018ApJ...860L...8S}. For the present case, by analyzing the periods of the associated flare we find that the periods of the QFP wave and the associated flare showed large difference. Therefore, we can disregard the mechanism related to the magnetic reconnection process. 3 and 5 minute oscillations are often detected respectively in coronal loops situated above sunspot and non-sunspot regions, and they are thought to be the leakage of photospheric oscillations into the corona \citep[e.g.,][]{2002A&A...387L..13D}. Due to the large difference of the period of the QFP wave to the 3 and 5 minute oscillations, it seems improbable that the observed QFP wave was driven by the leakage of photospheric oscillations into the corona. Considering the temporal and spatial relationships between the QFP wave and the eruption of the coronal jet, we propose that the most possible excitation mechanism of the present QFP wave was probably the dispersive evolution of disturbance caused by the interaction between the coronal jet and the loop system. Since the periods of the kink oscillation and the QFP wave are similar, this can lead to another alternative possible excitation mechanism for the QFP wave, i.e., the QFP wave might be excited by the interaction between the oscillating loop and the eruption jet plasma, in which the period of the QFP wave was modulated by the kink oscillation of the loop system.

For the excitation of the kink wave in the coronal loop, it was possibly excited by the impingement of the coronal jet, since statistical results suggested that most of kink waves are associated with lower coronal euptions \citep{2015A&A...577A...4Z}. \cite{2018ApJ...860...54O} recently found that the passage of  QFP waves in coronal loops can also excited kink oscillations of the loop system. For the present case, if the QFP wave originated from the flaring source region and propagated along the interconnecting loop system, the excitation of the kink oscillation of the loop system by the QFP wave was also possible. It is difficult for kink oscillations to excite QFP waves. Because kink oscillations are trapped in the loop for all values of the parallel wavenumber \citep[e.g.,][]{1983SoPh...88..179E}, the oscillations do not leak out and can not excite propagating fast waves in the external medium. In the future, theoretical and numerical simulation works are required for testing these new excitation mechanisms of the observed waves.

\section*{Acknowledgements}
The author thanks the observations provided by the {\em SDO}, and the referee's valuable suggestions and comments that highly improved the quality of the paper. This work is supported by the Natural Science Foundation of China (11773068,11633008,11403097,11533009), the Yunnan Science Foundation (2015FB191,2017FB006), the Specialized Research Fund for State Key Laboratories, and the Youth Innovation Promotion Association (2014047) of Chinese Academy of Sciences Sciences.

% Don't change these lines
\bsp	% typesetting comment
\label{lastpage}
\end{document}